\documentclass[letter]{aa} 
\usepackage{graphicx}
\usepackage{graphicx,lscape}
\usepackage{txfonts}
\usepackage{color}
\newcommand{\excs}{\extracolsep{\fill}}
\begin{document}

 \title{An accurate distance to 2M1207Ab
 \thanks{Based on observations collected at the European Southern
 Observatory, Chile (76.C-0543, 077.C-0112, 078.C-0158, 079.C-0229) and at Valinhos meridian circle.}\\}  

 \author{C. Ducourant \inst{1}
  \and R. Teixeira \inst{2,1}
   \and G. Chauvin \inst{3} \and G.Daigne
 \inst{1} \and J.F. Le Campion \inst{1} \and Inseok Song \inst{4} \and
 B.Zuckerman \inst{5}}

 \offprints{ducourant@obs.u-bordeaux1.fr}

 \institute{
 Observatoire Aquitain des Sciences de l'Univers, CNRS-UMR 5804, BP 89, 33270 Floirac, France.
           \and
 Instituto de Astronomia, Geof\'isica e Ci\^encias Atmosf\'ericas,
        Universidade de S\~ao Paulo,
        Rua do Mat\~ao, 1226 - Cidade Universit\'aria,
        05508-900 S\~ao Paulo - SP,
        Brasil. 
          \and
	  Laboratoire d'Astrophysique, Observatoire de Grenoble,
           414, Rue de la piscine, 38400 Saint-Martin d'H\`eres, France
          \and
	  Spitzer Science Center, IPAC/Caltech, MS 220-6, Pasadena, CA 
91125, USA
          \and
	  Department of Physics \& Astronomy and Center for Astrobiology, University of California, Los Angeles, Box 951562, CA 90095, USA} 
 \date{Received  / Accepted }

 \titlerunning {2M1207\,Ab distance and mass}


 
 
  \abstract 
  {In  April 2004 the  first image  was obtained  of a  planetary mass
  companion (now  known as 2M1207\,b) in orbit  around a self-luminous
  object  different from  our own  Sun (the  young brown  dwarf 2MASSW
  J1207334-393254, hereafter 2M1207\,A).   2M1207\,b probably formed via
  fragmentation and gravitational collapse, offering proof that such a
  mechanism can form bodies in the planetary mass regime. However, the
  predicted mass, luminosity, and radius of 2M1207\,b depend on its age,
  distance, and other observables such as effective temperature.}
  {To refine our knowledge of the physical properties of 2M1207\,b and
  its nature, we obtained an  accurate determination of the distance to
  the  2M1207\,A and  b system  by measurements  of  its trigonometric
  parallax at the milliarcsec level.}
  {With the  ESO NTT/SUSI2 telescope, in  2006 we began  a campaign of
  photometric   and   astrometric observations   to   measure   the
  trigonometric parallax of 2M1207\,A.}
  {An  accurate  distance  ($52.4\pm 1.1$  pc) to  2M1207A  was
  measured.   From  distance and  proper  motions  we derived  spatial
  velocities fully compatible with TWA membership. }
  {With  this  new  distance  estimate,  we  discuss  three  scenarios
  regarding  the  nature of  2M1207\,b:  (1)  a cool  ($1150\pm150$~K)
  companion   of    mass   $4\pm1$~M$_{\rm{Jup}}$,   (2)    a   warmer
  ($1600\pm100$~K)  and   heavier  ($8\pm2$~M$_{\rm{Jup}}$)  companion
  occulted  by   an  edge-on  circum-secondary  disk  or   (3)  a  hot
  protoplanet collision afterglow. }


   \maketitle
 

\section{Introduction}

 Since the  discovery of the enigmatic  classical T Tauri  star TW Hya
 isolated  from any dark  cloud (\cite{Rucinski}),  important progress
 has been made with the diagnostic selection and the identification of
 young stars near the Sun. In addition to the members of the TW Hydrae
 association (\cite{Kastner},  hereafter TWA), we  count nowadays more
 than 200 young ($<$100  Myr), nearby ($\le100$~pc) stars, gathered in
 different  clusters   and  co-moving  groups,  such   as  $\eta$
 Chamaleontis    (Mamajek   et    al.     1999),   $\beta$    Pictoris
 (\cite{zuck01}),  Tucana-Horologium   (\cite{torres00,  zuck00}),  AB
 Doradus (\cite{zuck04}) and the  most recent identified candidates of
 the SACY survey (Torres et  al. 2006). Their youth and proximity make
 these  stars ideal  sites for  study of  mechanisms of  planet, brown
 dwarf and  star formation. They  also represent favorable  niches for
 calibration   of   evolutionary   tracks   through   dynamical   mass
 measurements.   With   the   development   of  direct   imaging   and
 interferometric  techniques, the  circumstellar environment  of these
 young stars  is now probed down to  a few AU. The  number of resolved
 disks  ($\beta$ Pic,  Smith and Terrille 1984;  HR\,4796,  Schneider et
 al. 1999; TW Hya, Krist et al. 2000; AU Mic, Liu 2004), substellar companions
 (TWA5,  Lowrance  et  al.    1999;  HR7329,  Lowrance  et  al.  2000;
 GSC\,8048-00232,  Chauvin et  al. 2003  and AB\,Pic,  Chauvin  et al.
 2005) and   pre-main-sequence binaries   with   dynamically
 determined masses (HD\,98800, \cite{bode05}; TWA5, \cite{kono07}) continues 
 to increase.

 In the rush to discover companions with masses below that required to
 burn  deuterium,  Chauvin et  al.  (2004)  obtained  an image  of  an
 extrasolar companion of planetary mass.  This object was discovered
 near 2MASS\,J1207334-393254  (hereafter 2M1207A), a  brown dwarf (BD)
 member of the 8 Myr old TWA (Gizis 2002). 2M1207A is proposed to have
 a  near edge-on  accreting disk  (Gizis  2002, Mohanty  et al.  2003,
 Sterzik et al.  2004) and to drive a bipolar  resolved jet (Whelan et
 al.  2007).  This binary  system,  the  lightest  known to  drive  an
 outflow, offers new insights in  the study of mechanisms of formation
 and evolution of BDs, including their  disk and jet properties and physical 
 and
 atmospheric  characteristics of  objects as  light as  a  few Jupiter
 masses.

 \begin{table*}[t]
\caption{\label{resultats}Astrometric and Bessel photometric parameters for
  2M1207\,A measured with ESO NTT/Susi2.}
\begin{tabular*}{\textwidth}{@{\excs}lllllllll}
\hline\hline \noalign{\smallskip}
 $\alpha$  &  $\delta$  & $\mu_{\alpha abs}$  & $\mu_{\delta abs}$ &$\pi_{abs}$   & $d$     &   V    &  R    &    I    \\
 (2000)    &(2000)      &(mas/yr)         & (mas/yr)       & (mas)& (pc)& (mag)   & (mag)    &  (mag)\\  
\noalign{\smallskip}\hline \noalign{\smallskip}      
$12^{h} 07^{m} 33.460^{s}$ &$-39 \degr32\arcmin53.97\arcsec$ &-64.2$\pm$0.4&-22.6$\pm$0.4&19.1$\pm$0.4&52.4$\pm$1.1&  20.15 $\pm$0.19&18.08 $\pm$ 0.17&15.95 $\pm$ 0.13\\
 
\hline\noalign{\smallskip}
  \end{tabular*}
\end{table*}

 Although numerous  techniques have been devoted to  study this binary
 system (imaging,  spectroscopy, astrometry) at  different wavelengths
 (X-ray, UV,  visible, near-IR, mid-IR, radio),  its distance remained
 uncertain and not well constrained.  The initial distance estimate of
 $\sim70$~pc by Chauvin et al.   (2004) was improved by Mamajek (2005)
 who obtained an estimate  of $53\pm6$~pc based on the moving
 cluster method.  This method relies on the space motion determination
 for  the TWA  and the  proper motion  of 2M1207A.   The  space motion
 determination for  the TWA was based  on the four  members with known
 Hipparcos distance: TWA1,  TWA4 TWA9 and TWA11.  Song  et al.  (2006)
 suggested that significant uncertainties in the Hipparcos distance to
 TWA9 might  affect the cluster  distance estimation. They  proposed a
 new  distance estimate  of $59\pm7$~pc  based on  an  improved proper
 motion determination for 2M1207A scaled with the proper motion and the
 Hipparcos distance to HR\,4796\,A (TWA11).

 To  determine  a  firm  estimate  for  the  distance  to  2M1207  via
 measurement of its trignometric  parallax, since January 2006 we have
 conducted  astrometric and  photometric observations  at the  ESO NTT
 telescope.   At the same  time others  observational programme  were
 developed yielding results  similar to ours (Biller and Close 2007, Gizis at  al.  2007, Mamajek and Meyer 2007) and
 discussed below.   Our observations  are presented in  Section~2. The
 data  reduction and  analysis and  the result  of  this trigonometric
 parallax programme  are given  in Section~3.  Finally,  the physical
 properties and  different hypotheses for the nature  of 2M1207\,b are
 discussed in Section~4.
 
\section{Observations}

\subsection{Instrumental set-up and strategy}

 Astrometric  and photometric  (V, R,  I) observations  were performed
 with  the ESO  NTT telescope  equipped  with the  SUSI2 camera  which
 ensures  a nice  compromise between  the field  of  view $(5.5\arcmin
 \times 5.5\arcmin)$  and the pixel scale  $80.5$~mas/pixel assuring a
 reasonable number of field  stars (113) necessary to perform accurate
 astrometric  reductions.  Seven  sets of  data were  acquired between
 January  2006  and  May 2007  with  a  total  of fourteen  nights  of
 observation. All  astrometric observations  were realized in  the ESO
 I\#814 filter.  A calibration star with  known trigonometric parallax
 DEN 1048-3956 was also observed in order to validate our results.

\subsection{Differential colour refraction}

\begin{figure}[t]
\begin{center}
    \includegraphics*[angle=-90,width=\columnwidth]{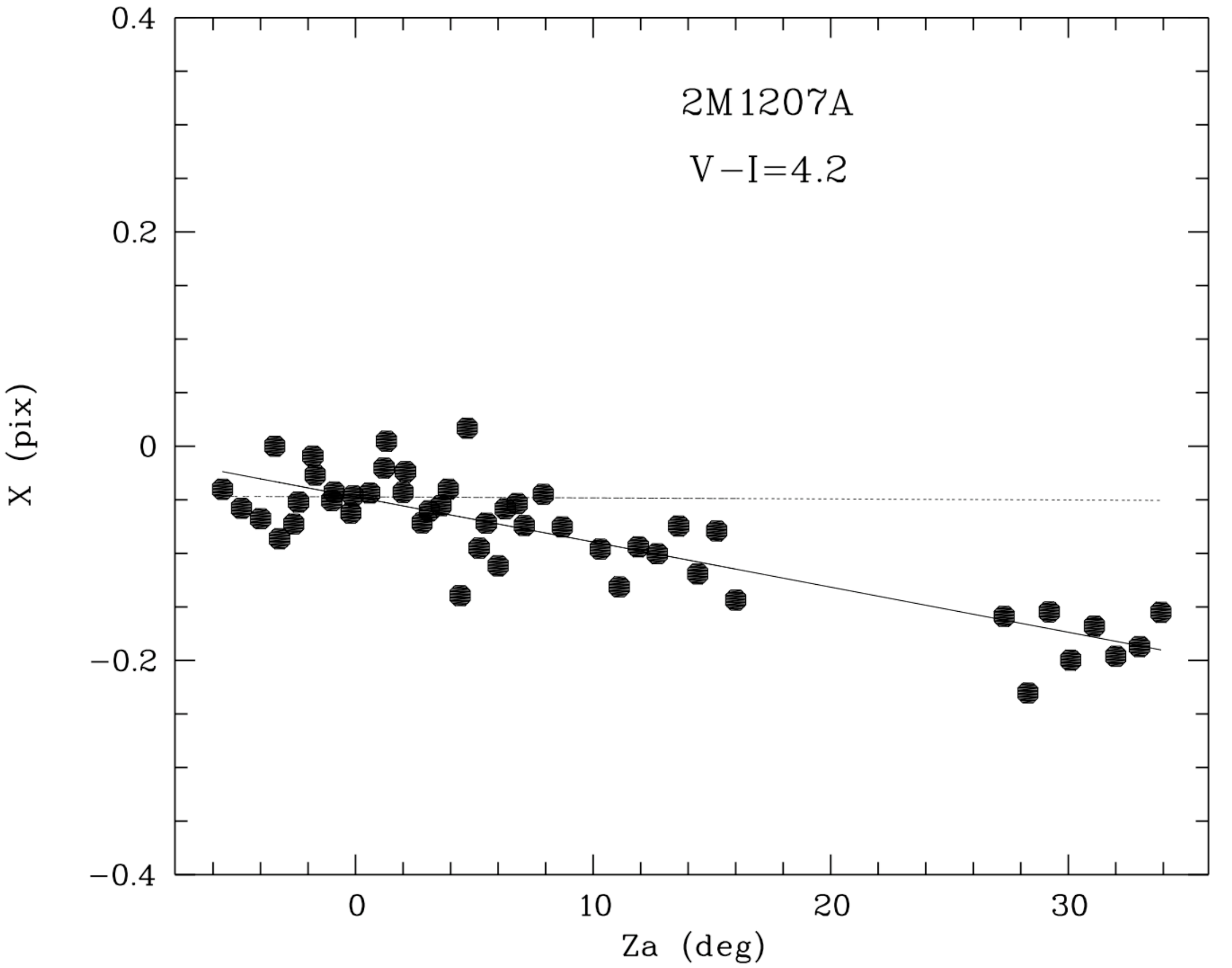}
    \caption{\label{dcr}Atmospheric refraction shift (X) in RA in the I
    filter as a function of projected tangent of the zenith distance in RA for
    2M1207. Horizontal line represents the mean atmospheric shift for
    reference background stars.}
\end{center}
\end{figure}

 Atmospheric refraction will affect differently our target brown dwarf
 and the  background reference stars (typically main--sequence  G or K
 stars) when observed through a given filter bandpass because of their
 difference  in  effective  wavelength.  This is  called  differential
 colour  refraction (DCR)  (Ducourant et  al. 2007)  which is  a major
 source of systematic errors in parallax programs. To minimize DCR all
 measurements were performed near  transit of the target (projected tangent of the zenith distance in R.A.
 $\le   20\degr$).  Following   Monet  et   al.  1992,   we  performed
 observations at small and large  hour angle during an observing night
 to  empirically calibrate  the difference  of refraction  between the
 target  and the  reference  stars.  We  present  this calibration  in
 Figure \ref{dcr}.  All measurements  of the target were corrected for
 DCR.
The difference of slope between 2M1207A and background stars is 
  $\Delta F=-0.004 \pm 0.001$ pix/deg. One can evaluate the accuracy 
  of DCR correction as $\sigma_{dcr}=\sigma_{\Delta F}<Z_{a}>=0.38$ mas 
  (with $<Z_{a}>=11.8\degr$, mean projected tangent of the zenith distance in right ascension of observations). 

\section{Data reduction and analysis}

\subsection{Astrometry}

\begin{figure}[t]
\begin{center}
    \includegraphics*[angle=-90,width=\columnwidth]{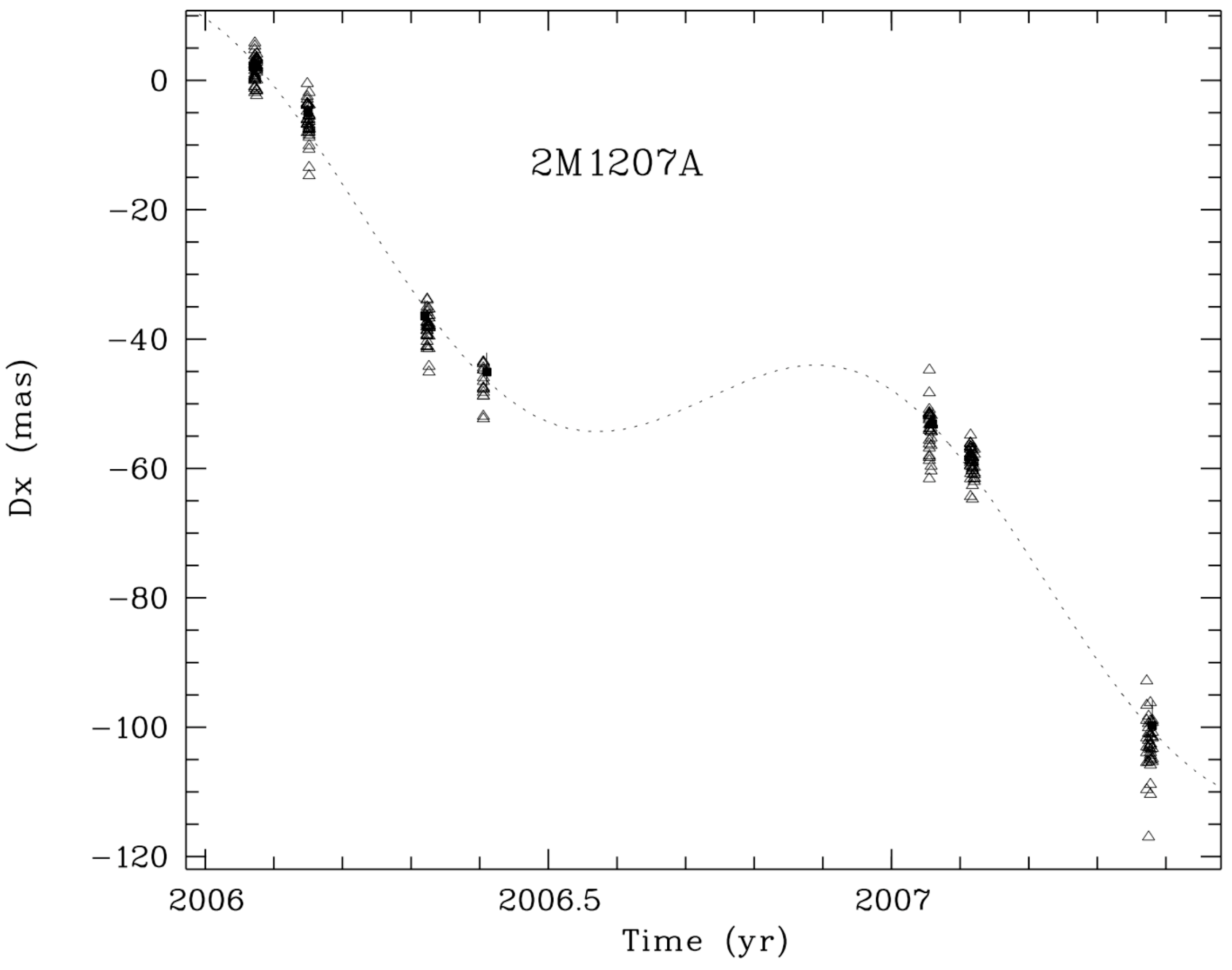}
    \caption{\label{pi}2M1207\,A   observations  in   right  ascention
    together with fitted path  along time : $\pi_{rel}=18.5 \pm 0.4$
    mas,   $\mu_{\alpha}^{*}=-57.5    \pm    0.4   $    mas/yr,
    $\mu_{\delta}=-22.5 \pm 0.4$ mas/yr.}
\end{center}
\end{figure}

 Images were measured using the {\tt DAOPHOT-II} package (Stetson
 1987) fitting a PSF to the images. The astrometric reduction of the
 whole dataset (261 images) was performed iteratively through a global
 central overlap procedure (see Ducourant et al. 2007) in order to
 determine simultaneously the position, the proper motion and the
 parallax of each object in the field. Each valid observation of each
 well-measured object in the field participates in the final solution
 (error on instrumental magnitude $\le 0.07$ (as given by DAOPHOT), Imag$\le 21.5$).  We present
 in Figure \ref{pi} the observations of 2M1207\,A together with the
 fitted path (relative parallax and proper motions).

\subsection{Conversion from relative to absolute parallax}

 As a  consequence of the least-square treatment,  the parallax and
 proper motion of the target is relative to reference stars (that are
 supposed to  reside at infinite  distance). The correction  from this
 relative  parallax  to  the  absolute  value  is  performed  using  a
 statistical  evaluation of the  distance of  the 113  reference stars
 ($13.5 \le I \le 21.5$), using the Besan\c con Galaxy model (Robin et
 al. 2003, 2004). This correction is $\Delta \pi=+0.58 \pm 0.01$ mas.
Same method is used to convert the relative proper motion of 
 2M1207A to absolute proper motion. 
 We derive the corrections: $\Delta \mu_{\alpha *}=-6.66\pm 0.04$ 
 mas/yr and $\Delta \mu_{\delta}=-0.02\pm0.02$ mas/yr.

 \subsection{Space motion}

 With our measured absolute parallax, together with a radial
 velocity of +11.2$\pm$2.0 $kms^{-1}$ from Mohanty et al.
(2003) and absolute proper motion, we calculate the Galactic space
 velocity (U,V,W) of 2M1207A as 
 (-7.9$\pm$0.8,-18.3$\pm$1.7,-3.5$\pm$0.8) $kms^{-1}$. The comparison
 of our results with the data from Reid (2003) (-10.0$\pm$2.6,
 -17.8$\pm$2.1, -4.6$\pm$1.1) confirms the compatibility of
 2M1207A motion with TWA membership.

\subsection{Control star}

 Our parallax  solution for DEN  1048-3956 is $\pi_{abs}=251.5\pm0.7$
 mas and  ($\mu_{\alpha}^{*},\mu_{\delta}$)=(-1170.0,-996.0)$\pm$ (0.6,0.6)
 mas/yr  which  is  in  good  agreement  with  Costa  et  al.  2005  (
 $\pi_{abs}=249.8\pm1.8$              mas,
 ($\mu_{\alpha}^{*},\mu_{\delta}$)=(-1175.3,-993.2)$\pm$       (2.2,2.2)
 mas/yr.

 We present in Table \ref{resultats} astrometric and photometric
 parameters for 2M1207A. 

\section{Discussion}
Since the discovery in 2004 of the planetary mass companion 2M1207\,b, several 
 groups have worked at its distance  determination
 (see table \ref{distances}). 
 Chauvin  et al. (2004)  first estimated  a distance  to and  mass for
 2M1207\,b of 70 pc and $5\pm2$~M$_{\rm{Jup}}$, and an associated effective
 temperature  of $1250\pm200$~K.   A low  signal-to-noise  spectrum in
 H-band enabled  them to suggest a  mid to late-L  dwarf spectral type
 for 2M1207\,b, supported by its  very red near infrared colors.  With
 a  revised  distance of  $53.3\pm6$~pc,  based  on  the moving  cluster
 method,   Mamajek  (2005)   re-estimated  the   mass   of  2M1207\,b.
 Converting  the $K_{s}$  absolute magnitude  into luminosity  using a
 bolometric   correction  appropriate  for   mid  and   late-L  dwarfs
 (Golimowski et al.  2004), based on his nearer  distance estimate, he
 derived  a   mass  of  $3-4$~M$_{\rm{Jup}}$  from  both   DUSTY  and  COND
 evolutionary models (Baraffe et al 2003).  With HST/NICMOS
 multi-band (0.9 to 1.6~$\mu$m)  photometry and a distance estimate of
 $59\pm7$~pc, Song  et al. (2006) derived a  mass of $5\pm3$~M$_{\rm{Jup}}$
 for 2M1207\,b,  due to its brighter flux than  expected from
 model predictions at  shorter wavelengths (for a given  mass and age)
 and to  the scatter in the  emergent flux predicted by  the DUSTY and
 COND03 evolutionary  models at wavelengths less  than 2.2~$\mu$m (and
 shown in Fig. 3).

\begin{table}[t]
\caption{\label{distances}Distance determinations for 2M1207A}
\begin{tabular*}{\columnwidth}{@{\excs}llll}
\hline\hline \noalign{\smallskip}

Authors      &        year &       Distance (pc) & Method \\
\hline \noalign{\smallskip}\noalign{\smallskip} 
Chauvin et al.     & 2004 & 70$\pm$20 & photometry \\
Mamajek & 2005 & 53.3$\pm$6.0 & moving cluster\\
Song et al.  & 2006 & 59 $\pm$7 & scale proper motion\\
Biller and Close & 2007 & 58.8$\pm$5.5 & trig. parallax\\
Gizis et al. & 2007 & 54$^{+3.2}_{-2.8}$& trig. parallax\\
Mamajek and Meyer & 2007 & 66$\pm$5 & moving cluster\\
This work  & 2007 & 52.4$\pm$1.1& trig. parallax \\
\noalign{\smallskip}\hline \noalign{\smallskip}      
 \end{tabular*}
\end{table}
   
\begin{table}[t]

\caption{\label{resb}DUSTY (Chabrier et al. 2000) and COND03 (Baraffe
 et al. 2003) predictions for the physical properties of 2M1207\,b for
(1) the hypothesis of a $1150\pm150$ K planetary mass companion (Chauvin et al. 2004) and,
 also,(2) a warmer $1600\pm100$~K and heavier $8\pm2$~M$_{\rm{Jup}}$ occulted
 planetary mass companion (Mohanty et al. 2007). Physical properties for (3) an
 alternative hypothesis proposed by Mamajek \& Meyer (2007) of a hot
 protoplanet collision afterglow are also given. In \textit{italic}
 are given the predicted fluxes, masses, effective temperatures, luminosities
 and surface gravities to be compared with the observed flux in $K_{s}$ and
 effective temperature of Chauvin et al. (2004) and Mohanty et al. (2007) respectively.} 

\begin{tabular*}{\columnwidth}{@{\excs}lllllll}
\hline\hline \noalign{\smallskip}

Hyp. & M$_{K_{s}}$ & $T_{\rm{eff}}$ & Mass & log(L/L$_\odot$) & log(g) \\
& (mag) & (K) & (M$_{\rm{Jup}}$) & (dex) & (dex) \\
\hline \noalign{\smallskip}\noalign{\smallskip} 
(1) & 13.33$\pm$0.13 & {\it 1150$\pm$150} & {\it 4$\pm$1} & {\it -4.5$\pm$0.2} & {\it 3.6$\pm$0.2} \\
(2)&\textit{11.2$\pm$0.3} & $1600\pm100$& \textit{8$\pm$2}     &                  \textit{-3.8$\pm$0.1}         &     \textit{3.9$\pm$0.1}               \\
(3)&$13.33\pm0.13$   &   $1600\pm100$    &   {\it   $\sim$0.25}   &   {\it-3.8$\pm$0.1} & {\it $\sim$3.0} \\
\noalign{\smallskip}\hline \noalign{\smallskip}      
 \end{tabular*}
\end{table}

 In addition to J-band photometry, Mohanty et al. (2007) obtained an 
 HK spectrum  of        2M1207\,b         at        low-resolution
 ($R_{\lambda}=100$).  Comparison with  synthetic spectra  DUSTY, COND
 and  SETTLE   yielded  an  effective   spectroscopic  temperature  of
 $1600\pm100$~K, leading  Mohanty et al.  (2007) to suggest a  mass of
 $8\pm2$~M$_{\rm{Jup}}$ for 2M1207\,b.   However, this mass and temperature
 are inconsistent  with that expected from model  predictions based on
 absolute  magnitudes  spanning $I$  to  $L\!'$-band photometry.  This
 discrepancy  is  explained  by  Mohanty  et  al.  (2007)  by  a  gray
 extinction of $\sim2.5$~mag between  0.9 and 3.8~$\mu$m caused by the
 occultation of  a circum-secondary  edge-on disk. This  hypothesis is
 illustrated in Fig.~3.

 Gizis et al (2007)  note that  the  hypothesis  of  a lighter  and  cooler
 2M1207\,b   cannot  be   ruled  out   by  the   current  photometric,
 spectroscopic and parallax  observations. Synthetic atmosphere models
 clearly encounter difficulties in describing faithfully the late-L to
 mid-T  dwarfs transition  ($\sim1400$~K for  field L/T  dwarfs). This
 transition corresponds to the process  of cloud clearing, that is, an
 intermediate  state between  two extreme  cases of  cool atmospheres:
 saturated in dust  (DUSTY) and where dust grains  have sunk below the
 photosphere (COND03). It is therefore probable that synthetic spectra
 fail for  the moment to properly model  the spectroscopic and
 physical characteristics  of young L  and T dwarfs.   Comparison of
 absolute fluxes  of 2M1207\,b with DUSTY and  CON03 model predictions
 for the age of the  TWA illustrates this possible intermediate state.
 In such  a scenario, absolute  magnitude and luminosity based  on the
 K$_s$-band photometry  indicate a mass  of $4\pm1$~M$_{\rm{Jup}}$ and
 an effective temperature of  $1150\pm150$~K. Such low temperature for
 a young  mid to  late-L dwarf would  corroborate the  observations of
 HD\,203030\,B  (Metchev \& Hillenbrand  2006) and  HN~Peg (\cite{luhman2007})
 indicating that the L/T  transition is possibly gravity dependent and
 appearing for temperatures as low as $\sim1200$~K.
\begin{figure}[t]
\begin{center}
    \includegraphics*[angle=0,width=\columnwidth,height=9.cm]{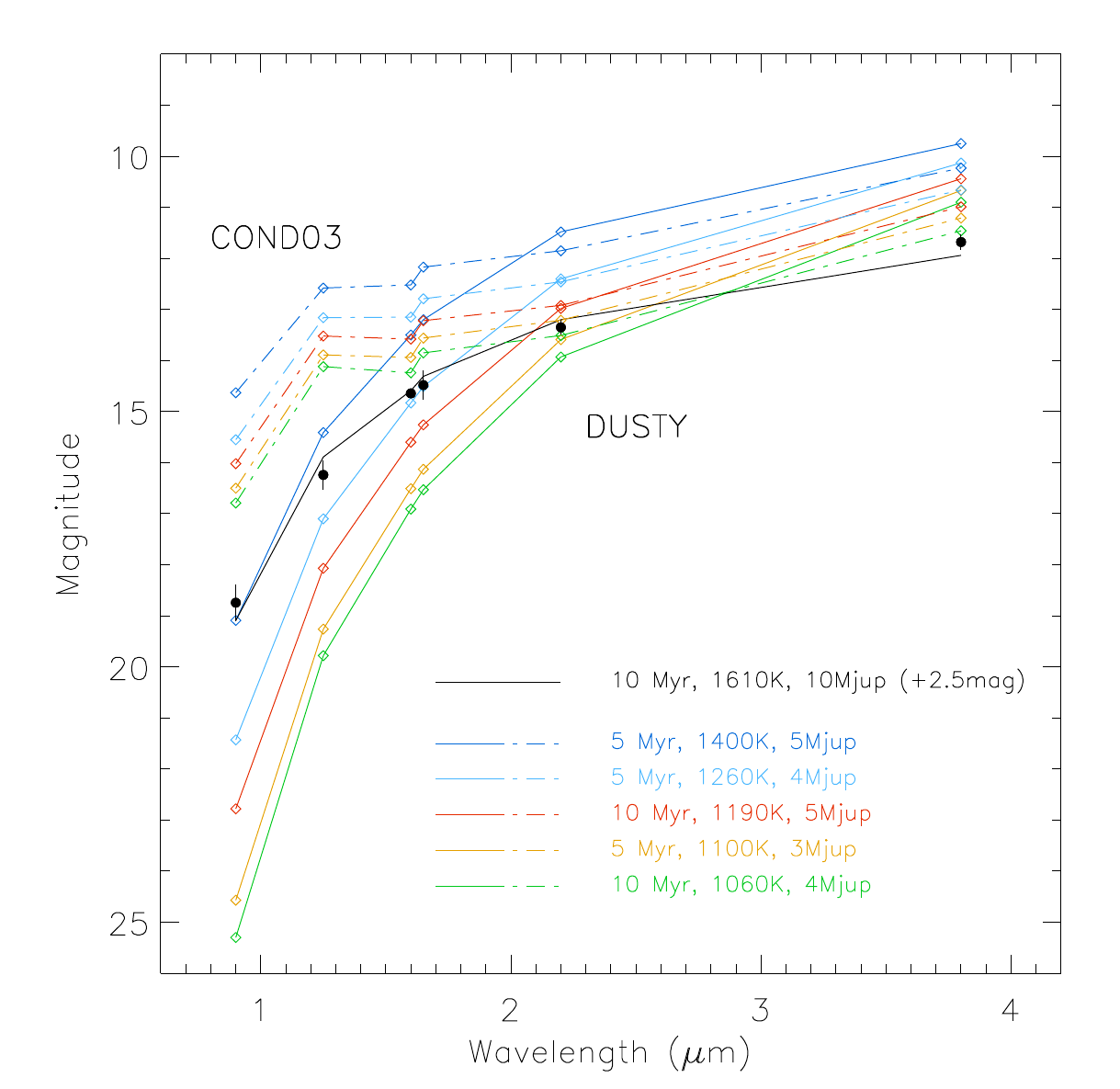}

    \caption{HST/NICMOS and VLT/NACO absolute magnitudes compared to
    predictions of the DUSTY (\textit{solid line}) and COND03
    (\textit{dash-dotted line}) evolutionary models at the age of the
    TWA and for a distance of 52.4$\pm$1.1~pc. VLT/NACO
    magnitudes have been converted into the CIT and Johnson-Glass
    system for direct comparisons with model predictions. Current
    observations are compatible with (1) a $4\pm1$~M$_{\rm{Jup}}$ planetary mass companion
    if DUSTY and COND03 atmospheric models fail currently to describe
    faithfully the atmosphere of 2M1207\,b or (2) a warmer $1600\pm100$~K
    and heavier $8\pm2$~M$_{\rm{Jup}}$ planetary mass companion occulted by a
    circum-secondary edge-on  disk (\textit{black solid  line}, with a
    $+2.5$~mag of grey extinction).} 

\end{center}
\end{figure}
 Our  current accurate parallax  measurement (52.4$\pm$1.1~pc)
 slightly changes the absolute  magnitudes and errors from ones
 given by  Mamajek (2005), Song et  al. (2006), Mohanty  et al. (2007)
 and Gizis et  al. (2007), but does not  modify their conclusions. The
 predicted   physical   properties   (mass,  luminosities,   effective
 temperature and gravity) are summarized in Table~3 for the two models
 described above.
 An alternative hypothesis has been proposed by
 Mamajek \&  Meyer (2007) to  explain the subluminosity  of 2M1207\,b.
 Their  explanation is that  the apparent  flux is  produced by  a hot
 protoplanet  collision  afterglow. Mamajek  \&  Meyer (2007) suggest
 several  ways  to  test this  hypothesis, for example a  lower
 surface gravity  ($\rm{log(g)}\sim3$) and a rich  metallicity for the
 protoplanetary  collision remnant. In  addition, potential observables
 such as polarized emission, infrared excess, resolved scattered light
 or a proposed $10~\mu$m  silicate absorption feature (predicted if 
 2M1207b is occulted by a circum-secondary  edge-on disk),  
 should help to clarify the nature of 2M1207\,b.

\section{Conclusion}

 Motivated by the need to have an accurate distance determination of the
 2M1207\,A and b system to better refine the properties of 2M1207b, we 
 measured the trigonometric parallax of the
 unresolved  system with a  precision better  than $2\%$.  This parallax puts
 2M1207\,A and b at 52.4$\pm$1.1~pc from our Sun and, along with our 
 accurately measured proper motion, lends substantial support to the notion
 that this binary is a member of the young TW Hydrae Association and, thus, 
 that the mass of 2M1207b lies clearly in the planetary mass range.  
\begin{acknowledgements}
      We would like to thank the staff of ESO-VLT and CFHT and Gilles Chabrier, Isabelle Baraffe and France Allard for providing the latest update of their evolutionary models. We also acknowledge partial financial support from the {\sl Programmes Nationaux de Plan\'etologie et de Physique Stellaire} (PNP \& PNPS) (in France),  the Brazilian Organism FAPESP and CAPES and French Organism COFECUB.
\end{acknowledgements}

\end{document}